\documentclass[twocolumn,aps,prb,showpacs,preprintnumbers,amsmath,amssymb]{revtex4}

\usepackage{epsfig}
\usepackage{graphicx}
\usepackage{dcolumn}
\usepackage{bm}

\begin{document}
\title{Phenomenological study of the electronic transport coefficients
of graphene}
\author{N.~M.~R. Peres$^1$, J.~M.~B. Lopes dos Santos$^2$, 
and T.~Stauber$^3$, }

\affiliation{$^1$Center of Physics and Department of
Physics, University of Minho, P-4710-057, Braga, Portugal}

\affiliation{$^2$ CFP and Departamento de F{\'\i}sica, Faculdade de Ci\^encias
Universidade de Porto, 4169-007 Porto, Portugal}

\affiliation{$^3$Instituto de Ciencia de Materiales de
Madrid. CSIC. Cantoblanco. E-28049 Madrid, Spain}



\date{\today}
\begin{abstract}
Using a semi-classical approach and input from  experiments on the 
conductivity of graphene, we determine the electronic density dependence
of the electronic transport coefficients -- conductivity, thermal
conductivity and thermopower -- of doped graphene. Also the electronic
density dependence of the optical conductivity is obtained. Finally
we show that the classical Hall effect (low field)
in graphene has the same form as for  the independent electron
case, characterized by a parabolic dispersion, as long
as the relaxation time is proportional to the momentum.
\end{abstract}
%
\maketitle
%
%


{\it Introduction.} 
Since the experimental measurement of an electric field effect
in graphene\cite{Nov04} and the observation of the odd integer
quantum Hall effect \cite{Nov05,Kim05}  the electronic
properties of graphene have been attracting a considerable attention
from the community.  Recent qualitative reviews
\cite{Nov07,Katsnelson07,peresworld} on the physics
of graphene give a brief account on both the experimental 
and theoretical status of the field.  

The measured conductivity of graphene has two distinct fingerprints:
a linear dependence upon the gate voltage of the DC conductivity,
and 
 a minimum  conductivity at the neutrality point of one electron
per carbon atom. The value of the conductivity minimum is of the order
of the quantum of conductance $e^2/h$. This minimum of conductivity is
a quantum mechanical effect which comes about as 
a consequence of disorder\cite{Peres06}. 
Disorder promotes a finite density of states at the Dirac point
which is responsible for the minimum of conductivity, $\sigma_{min}$. Unfortunately
there 
are
a number of different theoretical predictions  available for $\sigma_{min}$
 in the
literature
 \cite{Zigler07,Ludwig94,Ziegler98,Katsnelson05,Gusynin05,Tworzydlo06,Falkovsky06,Cserti06,Ziegler06}. Except for the 
numerical work of Nomura and  MacDonald\cite{Nomuraa} all the
calculations for the conductivity minimum are off by some numerical
factor from the experimentally measured value. 

Also interesting by  itself is the linear dependence of the
conductivity on the gate potential. Since the gate potential
depends linearly on the electronic density, $n$, one has a
conductivity $\sigma\propto n$. As shown  by Shon and  Ando\cite{Ando}
if the scatterers are short range one obtains a DC conductivity
that is independent of the electronic density, at odds with the experimental 
result. Nomura and  MacDonald\cite{Nomuraa,Nomurab} showed that
considering a scattering mechanics based on screened charged impurities
it is possible to obtain from a Boltzmann equation approach 
a conductivity varying linearly with the density, in agreement 
with the experimental result. In this brief report we address the problem
of finding the electronic density dependence of several transport
properties of graphene. Since the electronic density is easily
controlled by a gate voltage the expressions we derive here 
can certainly be tested experimentally. We therefore present in what follows
a calculation of 
the electronic density dependences of the DC conductivity, the thermal
conductivity and the thermopower. In addition, we study how the
electronic density enters in the optical conductivity. 
Finally,  we show that a certain form of the scattering rate
is necessary if the classical expression for the Hall effect is  to be
maintained, in agreement with the experiments.
A similar semi-classical approach was used by Falkovsky
\cite{Falkovsky} to study the temperature dependence of the Hall 
conductivity in graphene. The link between the Kubo-Streda
formulation and the semi-classical Boltzmann equation  approach was discussed,
in the context of massive Dirac fermions, 
by Sinitsyn {\it et al.}.\cite{Sinitsyn07}

{\it Boltzmann equation and relaxation times.}
The Boltzmann equation has the form \cite{Ziman}
\begin{equation}
-\bm v_{\bm k}\cdot\bm\nabla_{\bm r} f(\epsilon_{\bm k})
-e/\hbar(\bm E+\bm v_{\bm k}\times H)\cdot\nabla_{\bm k}f(\epsilon_{\bm k})
=-\left.\frac {\partial f_{\bm k}}{\partial t}
\right\vert_{scatt.}\,.
\label{boltzmann}
\end{equation}
The solution of the Boltzmann equation in its general form is difficult and one needs therefore to rely  upon some approximation. The first
step in the usual approximation scheme is to write the distribution
as $f(\epsilon_{\bm k}) = f^0(\epsilon_{\bm k})+g(\epsilon_{\bm k})$
where $f^0(\epsilon_{\bm k})$ is the steady state distribution function
and $g(\epsilon_{\bm k})$ is assumed to be small. Inserting this ansatz 
in Eq. (\ref{boltzmann}) and keeping
only terms that are linear in the external fields one obtains the 
linearized Boltzmann equation \cite{Ziman} which reads
\begin{eqnarray}
&-&\frac {\partial f^0(\epsilon_{\bm k})}{\partial \epsilon_{\bm k}}
\bm v_{\bm k}\cdot
\left[
\left(
-\frac{\epsilon_{\bm k}-\zeta}{T}
\right)\bm\nabla_{\bm r}T
+e\left(
\bm E-\frac 1 e\bm\nabla_{\bm r}\zeta
\right)
\right] =\nonumber\\
&-&\left.\frac {\partial f_{\bm k}}{\partial t}
\right\vert_{scatt.} +
{\bm v_{\bm k}\cdot \bm \nabla_{\bm r}}g_{\bm k}+\frac e {\hbar}
(\bm v_{\bm k}\times \bm H )\cdot \bm \nabla_{\bm k} g_{\bm k}\,.
\end{eqnarray}  
The second approximation has to do with the form of the
scattering term. The simplest approach  is to
introduce a relaxation time into the formalism.
This is done by considering the approximation
\begin{equation}
-\left.\frac {\partial f_{\bm k}}{\partial t}
\right\vert_{scatt.}\rightarrow\frac {g_{\bm k}}{\tau_{\bm k}}\,,
\end{equation} 
where $\tau_{\bm k}$ is the relaxation time, assumed to be
momentum dependent. Its momentum dependence 
will be determined
phenomenologically in such way that the dependence of the
conductivity upon the electronic density agrees with the
experimental data. 
The Boltzmann equation is certainly not
valid at the Dirac point, but since many experiments are done
at finite carrier density, controlled by an external
gate voltage,  we expect the Boltzmann equation to give
reliable results when an appropriate form for  $\tau_{\bm k}$
is used.

\begin{figure}
\begin{center}\includegraphics*[width=3.5cm]{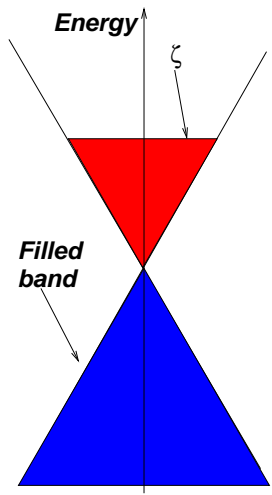}
\end{center}
\caption{\label{fig_cones}(Color online) 
Dirac cones in graphene, in a situation of finite electronic
density. The valence band is inert as long as the energy 
excitation processes are smaller than $\zeta$ (the Fermi energy). 
}
\end{figure}

Let us compute the Boltzmann relaxation time, $\tau_{\bm k}$, 
for two different
scattering potentials:(i) a delta function potential; (ii) a
screened Coulomb potential. The  relaxation time
$\tau_{\bm k}$ is defined as ($\bm k$ is the momentum)
\begin{equation}
\frac 1{\tau_{\bm k}}=N_iA \int d\,\theta
\int \frac{k'd\,k'}{(2\pi)^2}
S(\bm k,\bm k')(1-\cos\theta)\,,
\end{equation}
where $N_i$ is the total number of impurities in the sample,
$A$ is the area of the graphene sheet,
and the transition rate $S(\bm k,\bm k')$ is given by
\begin{equation}
S(\bm k,\bm k')=\frac {2\pi}{\hbar}\vert H_{\bm k',\bm k}\vert^2
\frac {1}{v_F\hbar}
\delta( k'- k)\,,
\end{equation}
where the $v_F\hbar k$ is the dispersion of Dirac fermions in graphene
and $H_{\bm k',\bm k}$ is defined as
\begin{equation}
H_{\bm k',\bm k} = \int d^2r\psi_{\bm k'}^\ast(\bm r)U_S(\bm r)
\psi_{\bm k}(\bm r)\,,
\end{equation}
with $U_S(\rm r)$  the scattering potential and $\psi_{\bm k}(\bm r)$
is the electronic spinor wave function of a clean graphene sheet. If the potential
is short range,\cite{Ando} of the 
form $U_S=v_0\delta(\bm r)$, the Boltzmann relaxation
 time turns out to be
\begin{equation}
\tau_{\bm k} =\frac {4\hbar^2v_F}{n_iv_0^2}\frac 1 k\,,
\end{equation}
where $n_i$ is the impurity concentration
per unit area.
On the other hand, if the potential is the screened Coulomb
potential, given by $U_S(\bm r)=eQe^{-r/L_D}/(4\pi\epsilon_0\epsilon r)$
for charged impurities of charge $Q$, the relaxation time is given by
\begin{eqnarray}
 \frac 1{\tau_{\bm k}} &=&\frac{u_0^2}{v_F\hbar^2k}
\left(1-\frac {\sqrt{1+4k^2L_D^2}-1}{2k^2L^2_D}
\right)
\,,
\end{eqnarray} 
where $u_0^2=n_iQ^2e^2/(16\epsilon_0^2\epsilon^2)$.
In the limit $L_D\rightarrow\infty$ one obtains
\begin{equation}
\tau_{\bm k} =\frac {v_F\hbar^2}{u_0^2}k\,.
\label{taucoulomb}
\end{equation}
As we argue below, the phenomenology of Dirac fermions
implies that the scattering in graphene must be of the form
(\ref{taucoulomb}). In what follows we explore the
consequences of this type of relaxation time.
 
At zero temperature the chemical potential $\zeta$ (Fermi energy)
is related to the density of charge carriers
(see Fig. \ref{fig_cones}) in the
conduction band by $n= {\pi}^{-1} \left( {\zeta}/{v_F\hbar}
\right)^2$. In what follows we give the expressions for the
transport coefficients in terms of $\zeta$.

{\it The DC conductivity.}
Within the relaxation time approximation the solution of the 
linearized Boltzmann equation when an electric field is applied to the
sample is
\begin{equation}
g_{\bm k} = -\frac {\partial f^0(\epsilon_{\bm k})}
{\partial \epsilon_{\bm k}}
e\tau_{\bm k}\bm v_{\bm k}\cdot\bm E\,,
\end{equation}
and the electric current reads 
\begin{equation}
\bm J=\frac {4}{A}\sum_{\bm k}e\bm v_{\bm k}g_{\bm k}\,.
\end{equation}
Since at low temperature the following relation
$-f^0(\epsilon_{\bm k})/\partial \epsilon_{\bm k}\rightarrow 
\delta (\zeta-v_F\hbar k)$ holds, one can easily see that assuming Eq. (\ref{taucoulomb})
where $k$ is measured relatively to the Dirac point,
the electronic conductivity turns out to be
\begin{equation}
\sigma_{xx} = 2 \frac {e^2}{h}\frac {\zeta^2}{u_0^2}=2 \frac {e^2}{h}\frac {\pi(\hbar v_F)^2}{u_0^2}n,
\label{sigmaxx}
\end{equation}
where $u_0$ is the strength of the scattering potential (with dimensions
of energy). The electronic  conductivity depends linearly on the
particle density, in agreement with the experimental data, 
\cite{Nov04,Nov05} and 
as  was first noted by Nomura and MacDonald 
in Ref. \cite{Nomurab}. The form used for $\tau_{\bm k}$ is thus imposed both by dimensional 
analysis and by phenomenology.  We note that for short range
scatterers\cite{Ando,Nomuraa}  $\tau_{\bm k}\propto k^{-1}$, leading to an 
DC conductivity that is independent of the density and therefore
in disagreement with the experiments.

Having settled the need for Eq. (\ref{taucoulomb})
we now proceed to determine the density dependence of the other
transport coefficients. We stress that the Coulomb potential
is one possible mechanism of producing a scattering rate of the
form (\ref{taucoulomb}) but we do not exclude that other mechanisms may exist.

{\it The optical conductivity.}
Here we want to obtain the electronic density dependence
of the optical conductivity of a doped graphene plane. Since the
Boltzmann approach does not include inter-band transitions, the 
expressions obtained below are only valid as long as $\hbar\omega\le \zeta\,$, where the above mentioned transitions are blocked by the Pauli principle.

Our aim is  to obtain  the response of the electronic system
to an external electric field of the form
\begin{equation}
\bm E =\bm E_0e^{i(\bm q \cdot \bm r -\omega t)}\,.
\end{equation}
The Boltzmann equation has, for this problem, the form
\begin{equation}
-\frac {\partial f^0(\epsilon_{\bm k})}
{\partial \epsilon_{\bm k}}
e\bm v_{\bm k}\cdot\bm E =
\frac {g_{\bm k}}{\tau_{\bm k}}  +
{\bm v_{\bm k}\cdot \bm \nabla_{\bm r}}g_{\bm k}+ \frac {\partial g_{\bm k}}{\partial t}\,.
\label{optical_boltzmann}
\end{equation}
The solution of the linearized Boltzmann equation (\ref{optical_boltzmann})
is well known \cite{Ziman}, reading
\begin{equation}
g_{\bm k}=-\frac {\partial f^0(\epsilon_{\bm k})}
{\partial \epsilon_{\bm k}}\Phi_{\bm q}(\omega,\bm k)
e^{i(\bm q\cdot \bm r -\omega t)}\,,
\end{equation}
with 
\begin{equation}
\Phi_{\bm q}(\omega,\bm k) = \frac {e\tau_{\bm k}\bm v_{\bm k}\cdot\bm E_0}{1-i\omega
  \tau_{\bm k}+i\tau_{\bm k}\bm q\cdot \bm v_{\bm k}}\,.
\end{equation}
The Fourier component, $\bm J(\omega,\bm q)$, of the current is given by
\begin{equation}
  \bm J(\omega,\bm q)= \frac {1}{\pi^2}
\int d^2k e\bm v_{\bm k}\Phi_{\bm q}(\omega,\bm k)\left(-\frac {\partial f^0(\epsilon_{\bm k})}
{\partial \epsilon_{\bm k}}\right),
\end{equation} 
leading in the long-wavelength limit to an optical conductivity  of the form
\begin{equation}
\sigma_{xx}(\omega)=
 2 \frac {e^2}{h}\frac {\zeta^2}{u_0^2}
\frac {1+i {\omega\hbar\zeta}/{u^2_0} }{1+
\left(
 {\omega\hbar\zeta}/{u^2_0}
\right)^2 }\;.
\label{optical}
\end{equation}
What should be stressed about Eq. (\ref{optical}) is its
density dependence. Since the electronic density is proportional
to $\zeta^2$ it should be possible to collapse the optical conductivity
curves for different densities in a representation of 
$\sigma_{xx}(\omega)/n$ versus $\omega\sqrt n$. This prediction
can easily be confirmed by reflectance measurements in graphene, for
frequencies below $\zeta$. Moreover, since the electronic density
induced by a gate voltage $V_g$ is proportional to the 
gate voltage, by the relation
$
n=\frac {\epsilon\epsilon_0}{te}V_g
$,
where $t$ is the thickness of the Silicon oxide substrate,
one obtains a collapse of the optical conductivity for different
gate voltages, by representing 
$\sigma_{xx}(\omega)/V_g$ versus $\omega\sqrt V_g$.

{\it Thermal conductivity and thermopower.}
Our next goal is to obtain the density dependence of the  thermal
conductivity and of the thermopower. For a discussion within the Kubo-formalism, see Ref. \cite{Sharapov1}.

In the presence of a temperature gradient in the sample,
the linearized Boltzmann equation has the form
\begin{equation}
-\frac {\partial f^0(\epsilon_{\bm k})}{\partial \epsilon_{\bm k}}
\bm v_{\bm k}\cdot
\left[
\left(
-\frac{\epsilon_{\bm k}-\zeta}{T}
\right)\bm\nabla_{\bm r}T
+e\bm E_{obsv.}
\right]=\frac {g_{\bm k}}{\tau_{\bm k}}\,,
\end{equation}
where the measured electric field is given
by $\bm E_{obsv.}=\bm E-\bm\nabla_{\bm r}\zeta/e$.
In this situation we have, in addition to the electric current,
a heat current (flux of heat per unit of area) given by
\begin{equation}
\bm U=\frac  {4}{A}\sum_{\bm k}\bm v_{\bm k}(\epsilon_{\bm k}-\zeta)
g_{\bm k}.
\end{equation}
Both the electric and the heat currents can be written as\cite{Ziman}
\begin{eqnarray}
\bm J &=& e^2\bm K_0\cdot \bm E_{obsv.}+\frac e T \bm K_1
\cdot(-\bm\nabla_{\bm r}T)
\nonumber\\
\bm U &=& e\bm K_1\cdot \bm E_{obsv.} + \frac 1 T \bm K_2
\cdot(-\bm\nabla_{\bm r}T)\,,
\end{eqnarray}
where $\bm K_i$, $i=0,1,2$ are second order tensors. In this problem
the tensors are diagonal, i.e. $\bm K_i=\bm 1 k_i$, and by a well established procedure \cite{Ziman} one obtains
\begin{eqnarray}
\label{k0}
  k_0&=& \frac 2 h \frac {\zeta^2}{u_0^2}\;,\\
\label{k1}
k_1&=& \frac 4 3 \frac {\pi^2}h(k_BT)^2\frac {\zeta}{u_0^2}\,,\\
\label{k2}
k_2&=& \frac 2 3 \frac {\pi^2}h(k_BT)^2\frac {\zeta^2}{u_0^2}\,.
\end{eqnarray} 
From the results (\ref{k0}), (\ref{k1}) and (\ref{k2}) it is easy
to derive both the thermal conductivity $\kappa$
and the thermopower $Q$. These are given by

\begin{equation}
\kappa = \frac 1 T
\left(
\frac 2 3 \frac {\pi^2}h(k_BT)^2\frac {\zeta^2}{u^2_0}
-\frac {8} 9\frac {\pi^4}h(k_BT)^4\frac 1 {u^2_0}
\right)
\end{equation}
and
\begin{equation}
Q = \frac 1{eT}\frac 2 3 \frac {\pi^2}{\zeta}(k_BT)^2\,.
\end{equation}
Again, what should be emphasized in these results
is the dependence of both
$\kappa$ and $Q$ on the particle density, which is different
from that of the usual two dimensional electron gas. Since it is 
experimentally feasible to control  the carrier density
in the graphene plane\cite{Nov04} 
it is possible to check experimentally the dependence of the
transport coefficients on the particle density.

{\it Hall effect.}
In what follows we prove that the scattering rate (\ref{taucoulomb}) leads to
a classical (low field)  Hall coefficient $R$
given by 
\begin{equation}
R=\frac 1 {en}\,,
\label{R}
\end{equation} 
in agreement with the experiments. For a discussion within the Kubo-formalism, see Ref. \cite{Sharapov2}.

The linearized Boltzmann equation in the presence of a static
magnetic field, perpendicular to the graphene plane $H=(0,0,B)$, is given by
\begin{equation}
\left( -\frac {\partial f(\epsilon_{\bm k})}{\partial \epsilon_{\bm k}}
\right) e\bm E\cdot \bm v_{\bm k}=\frac {g_{\bm k}}{\tau_{\bm k}}
+\frac e {\hbar} (\bm v_{\bm k}\times\bm H )\cdot \nabla_{\bm k}g_{\bm k}\,.
\label{Bhall}
\end{equation} 
The determination of $g_{\bm k}$ becomes quite simple by realizing
that the right-hand-side of Eq. (\ref{Bhall}) (after multiplication
by $\tau_{\bm k}$) is the first order expansion of 
\begin{equation}
g_{\bm k +\tau_{\bm k}({e}/{\hbar}) (\bm v_{\bm k}\times\bm H )}\,.
\label{gHall}
\end{equation}
Since $\bm v_{\bm k}=v_F\bm k/k$ and $\tau_{\bm k}$ is given by 
Eq. (\ref{taucoulomb}), Eq. (\ref{gHall}) can be rewritten as
$g_{\bm k+( {\alpha ev_F}/ {\hbar}) (\bm k\times\bm H )}\,$,
where $\alpha=v_F\hbar^2u^{-2}_0$. We now define a new momentum variable
$\bm k'$ given by
\begin{equation}
  \bm k'=\bm k+\frac {\alpha ev_F} {\hbar} (\bm k\times\bm H )\,.
\label{momentum}
\end{equation}
Solving Eq. (\ref{momentum}) for $\bm k$ as function of $\bm k'$
and making the left hand side of Eq. (\ref{Bhall}) equal to $g_{\bm k'}$,
one obtains the solution
\begin{equation}
g(\bm k) = 
\left( -\frac {\partial f(\epsilon_{\bm k})}{\partial \epsilon_{\bm k}}
\right)ev_F\alpha \bm k\cdot (\bm E -\alpha e v_F \bm H\times\bm E/\hbar)
/(1+a^2)\,,
\label{gbolt}
\end{equation} 
with $a=v_F^2\hbar^2/(l^2_Bu^2_0)$ and $l^2_B =\hbar/(eB)$.

Knowing the solution to $g(\bm k)$ the electric current is easily
computed giving
\begin{equation}
\bm J =\frac { \sigma_{xx}}{1+a^2}\left(
\bm E-a\hat z\times\bm E
\right)\,,
\end{equation}
where $\hat z$ is a unit vector along the $z$ direction.
The conductivity tensor $\bm \sigma$ is therefore given by
\begin{equation}
\bm \sigma =\frac {\sigma_{xx}} {1+a^2}
\left(
\begin{array}{cc}
 1& a\\
-a&1 \
\end{array}
\right)
\end{equation}
In the traditional Hall set up one has $J_y=0$ leading
to
\begin{equation}
J_{x}=\sigma_{xx}E_x\,,
\end{equation}
meaning that there is no magnetoresistance,
and
\begin{equation}
E_y=\frac {\pi v_F^2\hbar^2}{\zeta^2e}HJ_x
=
\frac 1 {ne}HJ_x\,,
\end{equation}
which produces  a $R$ Hall constant given by Eq. (\ref{R}).
This result is in agreement with the experimental findings
in the low field Hall effect \cite{Nov04,Nov05}. 
 
{\it Conclusions.}
In this brief report we derived,
using a semi-classical approach, the electronic density dependence
of the DC conductivity,  the optical conductivity, the thermal
conductivity, the thermopower and the classical (low field) Hall effect. 
Our proposed expressions
are based on  a phenomenological equation for the 
scattering rate of the Dirac electrons. Some of our findings
have already received experimental confirmation; the results
for the thermal properties and for the optical conductivity 
can be tested experimentally. The Boltzmann approach 
can not explain the universal conductivity value occurring
for $n\rightarrow 0$, that is, our approach breaks down at 
the neutrality point.

{\it Acknowledgments:}
The authors want to thank F. Guinea, Shan-Wen Tsai, and 
A. H. Castro Neto for useful comments and
many discussions. This work has been supported by MEC (Spain) through
Grant No. FIS2004-06490-C03-00, by the European Union, through
contract 12881 (NEST), and the Juan de la Cierva Program (MEC, Spain).
N.~M.~R.~P. and J.~M.~B.~L.~S. thank 
FCT under the grant PTDC/FIS/64404/2006.


\end{document}